\documentstyle[twocolumn,eqsecnum,epsfig,prl,aps,amsmath,amssymb,float]{revtex}
\begin{document}
\newfloat{figure}{ht}{aux}
\preprint{IUCM96-019}
\draft
\twocolumn[\hsize\textwidth\columnwidth\hsize\csname
@twocolumnfalse\endcsname
\title{Dynamical effects of phonons on soliton binding
in spin-Peierls systems}
\author{D.~Augier$^a$, D.~Poilblanc$^a$, E.~S\o rensen$^a$, and I. Affleck$^b$}
\address{$^a$Laboratoire de Physique Quantique \& 
UMR CNRS 5626, Universit\'e Paul Sabatier,
31062 Toulouse, France}  
\address{$^b$Department of Physics and Astronomy, and Canadian Institute for
Advanced Research,
University of British Columbia,\\
Vancouver, BC, V6T 1Z1, Canada}
\date{\today}
\maketitle
\begin{abstract}
The role of dynamical magneto-elastic coupling in spin-Peierls chains 
is investigated by numerical and analytical
techniques. We show that a Heisenberg 
spin chain coupled to dynamical optical phonons 
exhibits a transition towards a spontaneously dimerized state in a
wide range of parameter space.
The low energy excitations are characterized as solitons. 
No binding  between solitons occurs in the isolated 
spin-phonon chain
and the dynamical spin structure factor shows a broad magnon 
dispersion.
However, elastic interchain coupling can lead to the formation
of bound states. 

\end{abstract}
\vskip2pc]

Quasi-one-dimensional (quasi-1D) 
magnetic systems have recently received renewed experimental and theoretical 
attention with the observation of
the spin-Peierls transition in the two inorganic  
CuGeO$_3$~\cite{hase} and NaV$_2$O$_5$~\cite{isobe,ohama,weiden} compounds
which consist of weakly coupled spin-1/2 chains~\cite{horsch}.
While such a phenomenon was discovered in organic 
materials~\cite{organics}, the new inorganic compounds can be synthesised 
as relatively large single crystals allowing for new experimental
studies~\cite{hase,isobe,ohama,weiden,regnault,NMR,x-rays,ain,raman}.

The phase transition was inferred from 
an isotropic drop in the 
magnetic susceptibility~\cite{hase} at a transition temperature 
T$_{\mathrm{SP}}$ signalling a non-magnetic ground state (GS). 
The spin-Peierls transition is
characterized  by the opening of a spin gap, as has been observed 
by inelastic neutron scattering~\cite{regnault} (INS) and 
NMR spectroscopy~\cite{NMR,ohama}, accompanied by a 
distortion of the lattice observed in x-ray diffraction 
experiments~\cite{x-rays}.

The formation at low temperature of a non magnetic GS has raised the 
possibility of observing
topological magnetic excitations (solitons) as
proposed theoretically in spin-$1/2$ frustrated 
Heisenberg spin chains~\cite{haldane}
(the so called $J_1-J_2$ model, where $J_1$ and $J_2=\alpha J_1$ 
are the nearest-neighbor and next-nearest-neighbor 
exchange couplings, respectively).
At the Majumdar-Ghosh (MG) point, $\alpha=0.5$, the GS is 
doubly degenerate corresponding to 
the two simple dimer patterns (called A and B) 
obtained by a regular succession of 
disconnected singlet (valence) bonds.
In fact, for all $\alpha>\alpha_c$ ($\alpha_c\approx 
0.241$)~\cite{haldane,crit,castilla,white}
the GS was numerically shown to be dimerized.
The elementary excitations in this phase 
are easily depicted at the MG point: a soliton $s$ (antisoliton $\bar{s}$)
consists of an unpaired spin separating two dimer 
patterns A (B) and B (A)~\cite{shastry}.
These objects carry spin $\frac{1}{2}$ and can propagate thereby
acquiring a dispersion. 
A spin-$1$ magnon excitation can be viewed as the excitation of
a singlet bond into a triplet. However, in the $J_1-J_2$ model, 
such excitations decay into unbound soliton and antisoliton 
excitations.

The spin-Peierls materials are widely described in the
literature~\cite{boucher} in terms of a static 1D antiferromagnetic 
frustrated dimerized
Heisenberg chain~\cite{lieb} which includes, in addition to
the frustrating magnetic $J_2$ coupling, an explicit
dimerization $\delta$ of $J_1$~\cite{pytte}.
Interchain interactions are neglected in this description.
The values of $J_1$ and $J_2$ were estimated from a fit 
of the magnetic susceptibility at
high temperature and the dimerization $\delta$ by requiring the model to
have the experimental spin gap. Typical results such as
$J_1$=160~K, $\alpha=0.36$, $\delta=0.014$ were obtained for 
CuGeO$_3$~\cite{riera,castilla} and $J_1$=440~K, $\alpha=0$, $\delta=0.048$
for NaV$_2$O$_5$~\cite{weiden,augier1}. 
The $J_1-J_2-\delta$ model successfully predicts~\cite{fledder,poilblanc}
the experimentally observed $s\bar{s}$ 
bound state~\cite{regnault,ain}. Indeed, the static potential 
of strength $\delta$ lifts the degeneracy between
the two dimer A and B patterns. Therefore the energy cost of creating a
A--B--A defect scales approximately as the length of 
the B structure. A confining force proportional to 
$\delta$ and to the separation between $s$ and $\bar{s}$ 
comes then naturally out~\cite{affleck,uhrig,werner}. 
Besides the spin-$1$ magnons, this also
suggests the existence of singlet $s\bar{s}$ bound 
states~\cite{uhrig,nuthukumar,georges} 
which could be seen e.g. in Raman spectroscopy~\cite{raman}.

The static dimerized model has, nevertheless, some
drawbacks. It shows no spontaneous symmetry breaking (since
the dimerization is introduced {\it de facto} in the model)
and ignores phonon dynamics which are expected to 
be important when the phonon frequency and the 
energy scale of spin fluctuations become comparable. 

In this Letter we investigate a {\it dynamical} 
spin-phonon model which fully incorporates the phonon dynamics.
A weak coupling renormalization group (RG)
treatment~\cite{zimanyi} is used to argue that this model spontaneously
dimerizes even in the absence of interchain coupling (at $T=0$)
for large enough spin-phonon coupling at all phonon frequency.
Complementary numerical calculations are needed for arbitrary 
parameters and show that this
spontaneous dimerization (accompanied 
by a simultaneous opening of a gap in the spin excitation spectrum)
occurs in fact in a wide range of parameter space. 
The elementary excitations are characterized as 
solitons. 
We show the absence of $s\bar{s}$ binding for decoupled chains. However, 
in the presence of an elastic interchain coupling 
sharp spin-$1$ magnon excitations are recovered in the dynamical structure
factor. Therefore we believe that a correct description of real
materials must include the interchain coupling~\cite{uhrig2}.
 
The key ingredient of the model is the
magneto-elastic coupling; the exchange integral is 
{\it dynamically} modulated by 
the relative atomic displacements along the chains. 
Using independent phonon creation (destruction) operators
$b_i^{\dagger}$ ($b_i^{\phantom\dagger}$) on each bond (i,i+1)
the model reads~\cite{khomskii,affleck}, 
\begin{eqnarray}
\nonumber
H&=&J \sum_i [(1+g (b_i^{\phantom\dagger}
+b_i^\dagger))(\vec{S}_{i}.\vec{S}_{i+1}-\frac{1}{4})\\
\nonumber
&+&\alpha (\vec{S}_{i}.\vec{S}_{i+2}-\frac{1}{4})]
+H^0_{\mathrm{ph}}+H_\perp,
\end{eqnarray}
where g is the magneto-elastic coupling constant. 
Here, we assume dispersionless optical phonons of frequency $\Omega$ 
{\it i.e.}
$H^0_{\mathrm{ph}}=\Omega \sum_i(b_i^\dagger b_i^{\phantom\dagger}
+\frac{1}{2})$. Presumably a model
with on-site phonons will lead to similar results, however, the model
used here is probably the more relevant to CuGeO$_3$~\cite{geerstma}.
The interchain elastic
coupling $H_\perp$ 
will be discussed later.

The numerical results are based on Lanczos exact diagonalization 
(ED)~\cite{poilblanc_ed} of closed rings with up to L=14 sites supplemented by
finite size scaling analysis and a comparison with Bethe Ansatz 
exact results of the Heisenberg chain and Density  
Matrix Renormalization
Group (DMRG) calculations of the frustrated $J_1$--$J_2$ chain. 
A reliable treatment of the phonon dynamics 
is a difficult task.
Preliminary studies have been carried out by considering a single
$q=\pi$ phonon mode~\cite{augier2}. However, to investigate the {\it local}
$s\bar{s}$ interaction it becomes necessary to consider the 
complete multi-mode problem~\cite{sandvik}. 
We use here the variational treatment introduced by Fehrenbacher
based on phononic coherent states. 
Two phononic states per site are retained 
including the vacuum 
and a phononic coherent state. 
This approach is non perturbative, preserves the full dynamics of the
phonons, and becomes exact in the weak and strong coupling 
limits~\cite{fehrenbacher}. 

As a first step, we shall consider the case of an isolated 
spin-phonon chain 
($H_\perp=0$). A RG argument similar to the one proposed 
in Ref.~\cite{zimanyi}
can be used for this model for small $g^2\Omega/J$. Due to the SU(2)
symmetry and the absence of charge excitations there is only one
instantaneous interaction $h$, with a positive bare value, which is
marginally irrelevant~\cite{affleck}.
Integrating out the phonons
generates a retarded electron-electron interaction $\tilde{h}$
with a bare value $\tilde{h}(0)\approx -cg^2 J/\Omega$
(c is a positive constant). 
One can then renormalize both types of interactions
down to energies of the order of the phonon frequency.
Below that scale, all interactions become essentially instantaneous
leading to an effective Hamiltonian with new
instantaneous interactions obtained by adding the retarded
renormalized coupling $\tilde{h}(\Omega)$ to the
instantaneous one $h(\Omega)$. 
The RG equations are~\cite{zimanyi} $h'=h^2$ and 
$\tilde{h}'=\frac{3}{2}h\tilde{h}+\tilde{h}^2$,
involving the derivatives with respect to the logarithm of
the energy scale.
If the shift is large enough
to change the sign of the instantaneous coupling
({\it i.e.} $h(\Omega)+\tilde{h}(\Omega)<0$), the
system is in the spontaneously dimerized phase. This scenario always occurs
for a large enough coupling such that 
$cg^2>\Omega h(0)/(2J)$ in the limit $\Omega \ll J$.

\begin{figure}
\begin{center}
\epsfig{file=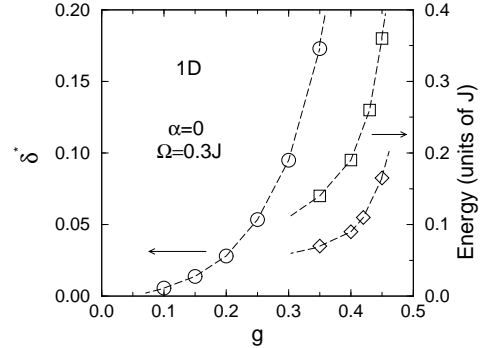,width=6cm}
\caption{Alternation in exchange integral $\delta^*$ ($\circ$), 
spin gap $\Delta^{01}$ ($\square$) and soliton minimum
energy $\Delta_S$ ($\lozenge$)
as a function
of the coupling g for $\alpha=0$ and frequency \hbox{$\Omega=0.3J$}.}
\label{para}
\end{center}
\end{figure}

We now show numerically the existence of a $T=0$
spontaneous symmetry breaking towards a dimerized phase
in a large region of parameter space.
The dimerized phase is signalled, 
in the thermodynamic limit, by
(i) the two fold degeneracy of the GS, (ii) the opening of a spin gap
and (iii) a lattice dimerization. 
The singlet-triplet spin gap $\Delta^{01}(L)$ as well as the separation
$\Delta^{00}(L)$ between the singlet GS in the $K=0$ and 
$K=\pi$ momentum
sectors were seen to scale accurately according to exponential 
laws~\cite{georges,augier1}.
The data strongly suggest $\Delta^{00}=0$, a first evidence for a
symmetry breaking in the GS. In addition, a finite extrapolated 
value of the spin gap $\Delta^{01}$ was obtained for 
a wide range of parameters.  As shown in Fig.~\ref{para}, $\Delta^{01}$
increases strongly with the magneto-elastic coupling $g$. 
Lower frequency phonons were found even more effective in opening 
the spin gap.

Then, to gain more insight we have 
computed the structural distortion associated to the broken symmetry. 
The GS correlation function $C_{\mathrm{latt}}(q)=\langle u_{-q}
u_q\rangle_0$ of the Fourier components
$u_q=L^{-1/2}\sum_i\exp(iqr_i) u_i$ of the relative lattice 
displacements ($u_i=b_i^{\phantom\dagger}+b_i^\dagger$) 
exhibits a divergence at momentum $q=\pi$ of the 
form $C_{\mathrm{latt}}(\pi)\propto L$.
The dimerization $\delta^*$ of the exchange integral is defined by
$\delta^{*2}=g^2\ \lim_{L\rightarrow\infty} \{L^{-1}C_{\mathrm{latt}}(\pi)\}$.
$\delta^*$ obtained from
a finite size scaling analysis 
is shown in Fig.~\ref{para} for $\alpha=0$ and $\Omega=0.3J$.
Its $g$ dependence follows closely the behavior of the spin gap.
According to the previous RG analysis, $\delta^*$ should be non zero
above a critical coupling, which is consistent with numerical results. 


\begin{figure}
\begin{center}
\epsfig{file=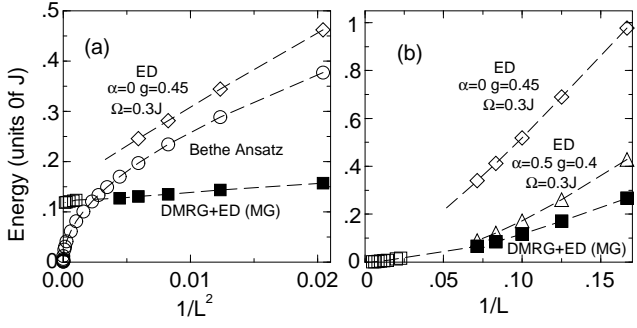,width=8.5cm}
\caption{(a) Minimum energy of the soliton 
$E_S$ as a function of the inverse square
of the chain length $1/L^2$ for $\alpha=0$, $g=0$ ($\circ$:
Bethe Ansatz), $\alpha=0.5$, g=0 ($\square$: DMRG, $\blacksquare$: ED) and
$\alpha=0$, $g=0.45$, $\Omega=0.3J$ ($\lozenge$, ED)
(b) Binding energy $E_B^{01}$ for $\alpha=0.5$, $g=0$ ($\square$:
DMRG, $\blacksquare$: ED)), $\alpha=0$, $g=0.45$, $\Omega=0.3J$ ($\lozenge$)
and $\alpha=0.5$, $g=0.4$, 
$\Omega=0.3J$ ($\triangle$) as a function of $1/L$.}
\label{soliton}
\end{center}
\end{figure}

We now turn to the characterization of the elementary magnetic excitations 
in the dimerized GS. If 
topological solitons exist in the spin-phonon model
such excitations should be created in pairs. 
However, finite chains with an odd number of sites cannot accommodate a
simple dimerized pattern and the GS of such a system is expected to
contain a single solitonic excitation. Therefore, the 
energy difference defined by
$E_S(k)=E_{\mathrm{odd}}^0(L,k)-E_{\mathrm{even}}^{0}(L)$, where 
$E_{\mathrm{odd}}^0(L,k)$ is the GS energy of the
chain of length $L=2p+1$ with momentum $k$ 
and $E_{\mathrm{even}}^0(L)$ is the absolute GS energy of
even chains interpolated at $L=2p+1$, can be considered as the 
excitation energy of a spin-$1/2$ topological defect propagating with 
a momentum $k$ along the chain. Note that, consequently
soliton energies  
cannot be computed in models including 
an explicit translation symmetry breaking, like e.g.
the dimerized Heisenberg chain.
To test the accuracy of this procedure, we have compared our results
for the spin-phonon model to the ones obtained for
the purely magnetic $J_1$--$J_2$ chain 
at the MG point where the spin correlation length is the shortest. 
In both cases, ED of periodic chains show a dispersion which can be 
parametrized 
as $E_S(k)\simeq\sqrt{\Delta_S^2+v_S^2(k\mp\pi/2)^2}$, where
$\Delta_S$ is the soliton gap and 
$v_S$ a characteristic velocity. 
The lowest excitation in a finite system is then obtained for the 
momentum which is closest to $\pi/2$~\cite{sca1}. 
For the spin-phonon model,
$\Delta_S$ appears to be finite for  
a wide region of parameter space~(see Fig.~\ref{soliton}(a)). Hence, the 
spin-$1/2$ excitation spectrum is massive indicating the existence
of solitons contrary to the case of the Heisenberg chain where 
spin-$1/2$ excitations (called spinons) are gapless~\cite{descloiseaux},
as explicitly shown in Fig.~\ref{soliton}(a) using the Bethe Ansatz solution
for $\alpha=0$~\cite{sca2}.
The soliton gap
should not depend on boundary conditions, and using
DMRG data for open chains~\cite{sca1}, we find
$\Delta_S\simeq0.1170(2)J$ at the MG point in the absence of a spin-phonon
coupling in good agreement with the estimate in Ref.~\cite{shastry}
of $\Delta_S\simeq0.125J$.

Spin-$1$ magnons can be considered as 
a $s\bar{s}$ combination in a triplet state. 
Consequently, the excitation triplet and {\it second}
singlet gaps are written 
as $\Delta^{ab}\simeq 2\Delta_S+E_B^{ab}$, 
where $E_B^{ab}$ accounts for a finite range spin-dependent
interaction between $s$
and $\bar{s}$. $E_B^{ab}$ is finite only if the 
$s\bar{s}$ combination forms a bound state, otherwise it vanishes and 
$s$ and $\bar{s}$ separate to infinity. The latter scenario
occurs for example in the spontaneously ($\alpha>\alpha_c$)
dimerized frustrated 
Heisenberg chain~\cite{affleck} where $\Delta^{00}=\Delta^{01}$.
However, in the case of spin-phonon models, a direct comparison between
$\Delta^{00}$ and $\Delta^{01}$ requires some caution: low 
energy spin-$0$ excitations of phononic character are likely to 
exist and are indistinguishable from the magnetic singlet $s\bar{s}$ 
excitations. The triplet binding energy $E_B^{01}=\Delta^{01}-2\Delta_S$ can
nevertheless be calculated 
on closed rings and ED results are shown in 
Fig.~\ref{soliton}(b).
The data for the spin-phonon model 
are very similar to the DMRG data 
on open chains at the MG point (also shown 
in Fig.~\ref{soliton}(b)) strongly suggesting
a vanishing binding energy for all parameters. 
This is consistent with the weak coupling 
RG treatment: one obtains an effective 
field theory similar to the one describing the pure
spin system (with renormalized coupling constants)
and no bound state is expected~\cite{affleck}.
We conclude that solitons and antisolitons are not bound in the 
1D spin-phonon model.

\begin{figure}
\begin{center}
\epsfig{file=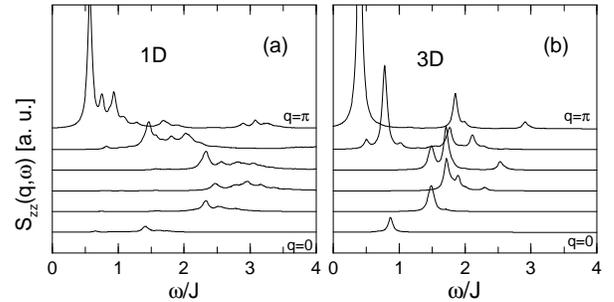,width=8cm}
\caption{Dynamical spin factor structure $S_{zz}(q,\omega)$ 
calculated at momenta 
$q=2\pi n/L$, $n=0,\cdots,L/2$ (L=12) 
(a) for $\alpha=0$, $g=0.3$, $\Omega=0.2J$ and no interchain coupling; (b)
for $\alpha=0$, $g=0.08$, $\Omega=0.2J$ and $\lambda=0.1J$. High 
energy peaks are due to finite size effects. A small broadening 
$\varepsilon=0.04J$ was used.}
\label{szzqw}
\end{center}
\end{figure}
Because of the decay of the $s\bar{s}$ pair, we do not expect 
any sharp $\delta$-function structure in the low frequency
spin structure factor. The latter which can be 
experimentally obtained by INS is defined as 
$S_{zz}(q,\omega)=\sum_n |\langle\Phi_n|S_z(q)|\Phi_0\rangle|^2
\delta(\omega-E_n+E_0),$
where $E_0$ ($E_n$) is (are) the energy(ies) of the GS $\Phi_0$ 
(triplet states $\Phi_n$) and $S_z(q)$ is the
Fourier transform of $S_z^i$.
Results for the spin-phonon model on a 12 site chain 
are shown in Fig.~\ref{szzqw}(a) for realistic parameters of NaV$_2$O$_5$
({\it i.e.} leading to a spin gap $\Delta^{01}\simeq0.2J$)
and are strikingly different from those
obtained for a static dimerization~\cite{augier1}. 
The main structure, although reminiscent 
of the well-known spinon dispersion of the Heisenberg 
chain ($\omega(q)=\frac{\pi}{2}J|\sin{q}|$)~\cite{descloiseaux}, 
is shifted towards higher energies and is much broader.
Furthermore, the relative weight of the low energy peak (e.g. at 
$q=\pi/2$) 
decreases for increasing system size.
Consistent with the previous analysis, there are no well
defined spin-1 magnons.

Next, we argue that the interchain coupling is crucial in order to
produce well defined magnon excitations as
experimentally observed. To test this scenario we have considered
an elastic interchain coupling of the form~\cite{affleck},
$H_\perp=K_\perp\sum_i
\sum_{\langle\gamma,\gamma'\rangle} u_i^\gamma u_i^{\gamma'}$,
where  
$\langle\gamma,\gamma'\rangle$ refers to adjacent  chains.
We shall treat this interchain elastic coupling
at the mean field level~\cite{khomskii,affleck}, while 
retaining the full dynamics of the 1D phonons. 
A given chain $\gamma$ will then be elastically coupled to the 
static deformation $\langle u_i^{\gamma'}\rangle_0=(-1)^i u_0 $ of
the $Z$ neighboring chains. Consequently we get an additional
{\it dynamical} term,
$H_{\perp,\mathrm{MF}}=\lambda \sum_i (-1)^i 
(b_i^{\phantom\dagger}+b_i^\dagger)$ where $\lambda=ZK_\perp u_0$ 
which explicitly doubles the unit cell. 

Turning on a small mean field interchain coupling $\lambda$
in the absence of frustration
could again favor one of the two lattice distortions and confine
solitons into pairs, in which case
a number of $s\bar{s}$ stable bound states proportional to 
$1/\lambda$ (in the limit $\lambda\ll J$) would be expected~\cite{affleck}.
Although the previous analysis on
soliton binding cannot be done anymore, we give further
arguments in favor of soliton binding based on the study of the
dynamical structure factor. This one is shown on Fig.~\ref{szzqw}(b) for 
parameters relevant to NaV$_2$O$_5$  
with a non-zero value of $\lambda$.
Qualitatively, the low energy magnon structure 
is now much better defined 
and the magnon dispersion is clearly apparent.
The maximum of the dispersion occurs at $\omega\approx1.7J$,
an energy close to the exact value $\frac{\pi}{2}J$ of the 
Heisenberg chain~\cite{descloiseaux}.
At momentum $q=\pi/2$ where finite size effects
are shown to be quite small~\cite{poilblanc,augier1},
the average frequency $\langle \omega \rangle$ and the
width $\Delta\omega$ of the structure depend strongly on the presence of an 
interchain coupling. 
The structure clearly gets narrower when the interchain 
coupling increases. 
Furthermore, when $\lambda\ne 0$ the relative weight of the lowest energy peak
increases with the length of the chain,
contrary to the case of the isolated chain.

We thank IDRIS (Orsay) for allocation of CPU time on the C94 and C98
CRAY supercomputers. The research of IA is supported in part by NSERC of
Canada.


\begin{references}
\bibitem{sca1} If $\Delta_S\ne0$, one expects $1/L^2$ corrections in
the energy.
\end{references}
\end{document}